\documentclass[12pt]{article}

\usepackage[totalheight = 23cm, totalwidth = 17cm]{geometry}
\usepackage{amssymb,amsmath,amsfonts,amsbsy,graphicx}
\usepackage{bm}
\usepackage{color}
\usepackage{subfigure}
\usepackage{mathtools}

\DeclarePairedDelimiter\floor{\lfloor}{\rfloor}

\newcommand{\mpl}{m_{\rm Pl}}

\newcommand{\be}{\begin{equation}}
\newcommand{\ee}{\end{equation}}
\newcommand{\bea}{\begin{eqnarray}}
\newcommand{\eea}{\end{eqnarray}}

\def\half{{1\over 2}}
\def\({\left(}
\def\){\right)}
\def\[{\left[}
\def\]{\right]}

\def\m{\begin{eqnarray}}
\def\n{\end{eqnarray}}

\begin{document}

\begin{titlepage}

\begin{center}

\rightline{???}

\vskip 1cm

%\LARGE{\bf Modulations of the scalar power spectrum from a heavy field with a monomial potential}

\LARGE{\bf Power-law modulation of the scalar power spectrum from a heavy field with a monomial potential}

\vskip 1cm

\large{Qing-Guo Huang$^{a,b}$~and~
Shi Pi$^{a}$}

\vskip 1.2cm

\small{\it
$^{a}$ Key Laboratory of Theoretical Physics, Institute of Theoretical Physics,\\ Chinese Academy of Sciences, Beijing 100190, China
\\
$^{b}$ School of Physical Sciences, University of Chinese Academy of Sciences, \\
No. 19A Yuquan Road, Beijing 100049, China
\\
\vspace{2em}
E-mail: huangqg@itp.ac.cn, spi@itp.ac.cn
}

\vskip 1.2cm

\end{center}

\begin{abstract}

The effects of heavy fields modulate the scalar power spectrum during inflation. We analytically calculate the modulations of the scalar power spectrum from a heavy field with a separable monomial potential, i.e. $V(\phi)\sim \phi^n$. In general the modulation is characterized by a power-law oscillation which is reduced to the logarithmic oscillation in the case of $n=2$. 

%Features from the primordial power spectrum of the curvature perturbation can be used to detect the heavy fields during inflation. We study the features from a two-field inflation with a seperable potential, of which the potential of the heavy field has a power-law of an arbitrary power $n$. We calculate analytically under slow-roll condition and massive limit, and find some new features in the power spectrum which can be used to fix the power of the heavy sector in inflation if such a signal is detected in the future. 

\end{abstract}

\end{titlepage}

\setcounter{page}{0}
\newpage
\setcounter{page}{1}

\section{introduction}\label{sec:intro}

Inflation is a leading paradigm for describing the physics in the early universe~\cite{inflation}. Besides solving the flatness, horizon, and abundant topological relics problems in the hot big bang cosmology, inflation can provide a natural mechanism to generate the primordial density perturbations seeded the CMB temperature anisotropies and the formation of large-scale structures observed today~\cite{perturbation}. The primordial power spectrum of the curvature perturbation, which originated from the quantum fluctuations deep inside the horizon during inflation, has been confirmed to be of order $10^{-5}$ and nearly scale-invariant~\cite{Ade:2015lrj}. 

The inflation is supposed to be driven by a canonical scalar field with a flat enough potential in the simplest version. The energy scale of inflation occurred in the very early universe is likely to be extremely high, such as the GUT scale ($\sim 10^{16}$ GeV). Up to now the Higgs is the unique fundamental scalar field which was confirmed by experiments~\cite{Aad:2012tfa}. However there is a mass hierarchy problem for Higgs field in the standard model of particle physics. Besides, identifying Higgs boson as the scalar field that drove inflation will require a non-minimal coupling to gravity, which keeps the integrity of the standard model of particle physics, yet extend the ``standard model'' in gravity. In general, ones believe that a new theory beyond the standard model is needed, and the inflation should be described in the framework of such a theory. For instance, as is invoked by the low energy realization of string theory or supergravity, multiple number of new degrees of freedom become relevant to the inflation. For the multi-field case when the extra fields have masses much smaller than the Hubble parameter $H$ during inflation, it is thoroughly studied and discussed, for instance, in \cite{twolightfield}. The quantum fluctuations from these fields are generally coupled with each other even if there is no such direct coupling in their potential. Therefore the curvature perturbation will not be conserved even after the horizon-crossing, which can bring some extra $k$-dependence in the final result of primordial power spectrum

In this paper, we switch to another limit in which there is a heavy field with effective mass larger than the Hubble parameter during inflation. In such a two-field model, if the heavy field is excited, either by a sudden kick or by some special shapes of the potential like those in~\cite{kick,shape}, it will affect the evolution of curvature perturbation, and usually some oscillatory feature will appear in the power spectrum as well as the bispectrum. The reason is that the underdamped solution for the heavy field will interact with the inflaton field mainly via the dynamical background. Here we we focus on the case with a heavy field whose potential takes the general form 
\m\label{intro2}
V(\phi)={1\over n}m^{4-n}\phi^n, 
\n 
where $n(\geq 2)$ is a constant, but not necessarily an integer. Such a potential might emerge in string or supergravity theories. See, for example some concrete models in \cite{Baumann:2014nda}. For $n=2$, the scalar field $\phi$ is always heavy if $m\gg H$; for $n> 2$, we can define an effective mass by $m_\text{eff}(\phi)\equiv \sqrt{d^2V(\phi)/d\phi^2}$, and a ``heavy'' field with potential \eqref{intro2} means $m^{4-n}\phi_\text{amp}^{n-2} \gg H^2$ where $\phi_\text{amp}$ is the oscillation amplitude of the $\phi$ field when it oscillates around its local minimum $\phi=0$. We analytically calculate the corrections from the oscillation of the heavy field to the primordial power spectrum of curvature perturbation, and find that a power-law modulation of the power spectrum generically shows up. For example, we get a concise expression for an integer $n>4$ as follows 
%\textcolor{red}{(The prefactors of $k$ and $\mu$ are not universal for $n=2$)}
\be\label{intro3}
\frac{\Delta\mathcal{P_R}}{\mathcal{P}_\mathcal{R}^{(0)}}\sim
\({\frac{k}{\mu k_*}}\)^{\frac52s-3}\mu^{-3/2} \sin\left[{C(s)\over s} \mu \left(\frac{k}{\mu k_\ast}\right)^s+\text{phase}\right],
\ee
where 
\be
s=\frac{3n-6}{2n-8}, 
\label{sn}
\ee
$C(s)$ is a function of $s$, $\mu\sim m_\text{eff}(\phi_\ast)/H$ is the dimensionless effective mass at $\phi_\ast$.
%, and the power of $k$ in the argument of the sinusoidal function $s$ is related to $n$ by
%For $n=3$, the $k$-dependent envelope of the sinusoidal function should be modified to $k^{-39/4}$, and there is no oscillating modulation for $n=4$. 
For $n=4$ the oscillatory feature is highly suppressed, while the modulation for $2\leq n<4$ is almost the same as that for $n>4$ except the scaling laws of the envelop: $\mu^{1/2} \({k/(\mu k_*)}\)^{-3}$ for $n=2$ and $\mu^{1/2} \({k/ (\mu k_*)}\)^{-39/4}$ for $n=3$. Note that the result for $n=2$ can be recovered by taking the limit of $s\rightarrow 0$, and the power-law modulation with $k^0$ is reduced to the logarithmic oscillation modulation, like that in~\cite{Polarski:1992dq,standardclock},
\be\label{intro1}
\frac{\Delta\mathcal{P}_\mathcal{R}}{\mathcal{P}_\mathcal{R}^{(0)}}\sim \({k\over \mu k_*}\)^{-3}\sin\left(2\mu\ln {k\over k_*} +\text{phase} \right).
\ee 
The observational signals on the CMB anisotropies from the features in Eq.~\eqref{intro3} has been already analysed in~\cite{Chen:2012ja}, based on a completely different physical picture.

This paper is organized as followed. In section \ref{sec:bgd} we study the background evolution of the heavy field $\phi$ with an arbitrary monomial potential of $m^{4-n}\phi^n$. After semi-analytically solving the equation of motion for $\phi$, we calculate the corrections to the Hubble  and slow-roll parameters. In section \ref{sec:pert} we calculate the corrections to the equation of motion for the curvature perturbation during inflation, and compute the primordial power spectrum of curvature perturbation. For an arbitrary $n$, the corrections to the primordial power spectrum can be expressed by some integrals, while for an integer $n$ we can get the analytic expression by the stationary phase method. Summary and discussion are given in section \ref{sec:con}.

\section{Background evolution and corrections to the slow-roll parameters}\label{sec:bgd}

In this paper we focus on the simplest case in which the heavy field does not directly couple to the inflaton field $\phi_{\rm inf}$. The Friedmann equation takes the form  
\be\label{Friedmann}
3\mpl^2H^2=\frac12\dot\phi_{\rm inf}^2+V_\text{sr}(\phi_{\rm inf})+\frac12\dot\phi^2+\frac1nm^{4-n}\phi^n,
\ee
where $\mpl=1/\sqrt{8\pi G}$ is the reduced Planck scale, $V_\text{sr}(\phi_{\rm inf})$ is the potential of inflaton field $\phi_{\rm inf}$ which provides slow-roll inflation in $\phi_\text{inf}$-direction. Here the power index $n (\geq 2)$ for the potential of $\phi$ field may not be an integer if we take, for instance, brane inflation into account. For $n=4$, the general potential of the heavy field should have taken the form of $\lambda\phi^4$, but $\lambda$ can be absorbed into $\phi$ by a redefinition $\phi\rightarrow\lambda^{-1/2}\phi$. Here we suppose that the energy density of the heavy field $\phi$ is subdominant, and the Hubble parameter $H$ is governed by the potential energy of inflaton field, namely 
\m
H^2\simeq {V_\text{sr}(\phi_{\rm inf})\over 3\mpl^2},
\n
which is roughly a constant during inflation.

The heavy sector might be excited from its local minimum either by a sudden kick caused by a turn~\cite{kick}, by a non-trivial initial condition~\cite{Polarski:1992dq}, or by some specific potential shape~\cite{shape}. Let us leave aside the mechanisms of the excitation, but only focus on an initial deviation from the vacuum denoted by a non-zero initial value $\phi_*$ at time $t_*$. Then the heavy field $\phi$ start oscillating around its local minimum. 
It is well-known that the energy density of a heavy scalar field which is rapidly oscillating around the local minimum of $V(\phi)\sim \phi^n$ goes like, in \cite{Turner:1983he}, 
\m
\rho_\phi\sim a^{-6n/(n+2)}
\n 
which implies that the amplitude of the oscillation of such a heavy field decreases as 
\m\label{phiamp}
\phi_\text{amp}=\phi_* \left(\frac{a}{a_*}\right)^{-{6\over n+2}}, 
\n
%here $a(t_*)=1$. 
It is convenient to introduce a dimensionless parameter to describe the fraction of the energy density of heavy field in the total energy budget during inflation 
\be
\Omega_\phi\equiv\frac{m^{4-n}\phi_\text{amp}^n}{3n\mpl^2H^2}
\ee
which has the same scaling behavior as $\rho_\phi$, namely 
\m
\Omega_\phi=\Omega_{\phi*} a^{-6n/(n+2)}, 
\n 
where 
\m 
\Omega_{\phi*}={m^{4-n}\phi_*^n\over 3n\mpl^2H^2} 
\n 
denotes the initial value of the energy fraction of field $\phi$.

For $n=2$, the mass of $\phi$ is a constant $m$, and $\phi$ is called a ``heavy field'' if $m\gtrsim H$. For general $n$, we define the effective mass of $\phi$ as the square root of the second derivative of $V(\phi)$ with respect to $\phi$,
\m\label{def:mass}
m_{\rm eff}(\phi)\equiv \[{d^2V(\phi)\over d\phi^2}\]^\half=\sqrt{(n-1)m^{4-n}\phi^{n-2}}
\n
which depends on the value of $\phi$. Here the scalar field $\phi$ can be taken as a heavy field if $m_{\rm eff}(\phi_\text{amp})> H$. For simplicity,  we can define a new dimensionless mass parameter $\mu$ by 
\be\label{def:mu}
\mu\equiv%\frac{V_{\phi\phi}}{V}=(n-1)
\frac{\sqrt{m^{4-n}\phi_*^{n-2}}}{H}. 
\ee 
For $n=2$, $\mu$ is nothing but $m/H$. For $n>2$, it is $m_\text{eff}/H$ up to a numerical factor at the beginning of oscillation. %describes the initial value of $m_\text{eff}$. 
So for a heavy $\phi$ field, we need $m_\text{eff}(\phi_\text{amp})\gtrsim H$, which, together with \eqref{def:mass} and \eqref{phiamp}, gives the e-folding number for $\phi$ staying heavy:
%The effective mass of scalar field $\phi$ for $n>2$ becomes comparable to the Hubble parameter during inflation when 
%\m
%\frac{a}{a_*}\simeq \mu^{n+2\over 3(n-2)}. 
%\n
%Roughly speaking, the e-folding number corresponding to the heavy scalar field $\phi$ reads 
\m\label{efoldheavy}
N_e^\text{heavy}\lesssim{n+2\over 3(n-2)}\ln \left[\sqrt{n-1}\mu\right],\;\;\;\;\text{for}~n>2.
\n
If the initial effective mass of $\phi$ is much larger than the Hubble parameter $H$, i.e. $\mu\gg1$, a few e-foldings for $\phi$ being heavy is expected. 
%%%%%%%%%%%%%%%%%%%%%%%%%%%%%%%%%%%%%%%%%%%%%%%%
%%%%%%%%                       Comments on n<2                               %%%%%%%%%%%%%%
%%%%%%%%%%%%%%%%%%%%%%%%%%%%%%%%%%%%%%%%%%%%%%%%
\iffalse
We can also see that for $n\leq2$, %$N_e^\text{heavy}\rightarrow\infty$, which means 
$\phi$ will always be heavy if it is so initially. 
%On the other hand, for $n<2$, $N_e^\text{heavy}<0$, and $\phi$ will never be heavy even if $\mu\gg1$. 
In this paper we will mainly focus on the status when $\phi$ stays heavy. This means $1<n\leq2$, or $n>2$ with \eqref{efoldheavy} is valid, both under the condition $\mu\gg1$. 
\fi
%%%%%%%%%%%%%%%%%%%%%%%%%%%%%%%%%%%%%%%%%%%%%%%%
%%%%%%%%                       Comments on n<2                               %%%%%%%%%%%%%%
%%%%%%%%%%%%%%%%%%%%%%%%%%%%%%%%%%%%%%%%%%%%%%%%
A useful relation from combing the definition of $\Omega_{\phi*}$ and $\mu$ gives
\be
\Omega_{\phi*}=\frac{\mu^2}{3n}\left(\frac{\phi_\ast}{\mpl}\right)^2.
\ee
We can either keep $\Omega_{\phi*}$ or $\phi_\ast/\mpl$ in our final result. The former one has better physical meaning, while the latter one helps us to understand the correct $\mu$-dependence when compared with $n=2$ case. The fact that the energy density of the heavy field is subdominant implies $\phi_*\ll \mpl/\mu$ where $\mu\gg 1$.

Now we turn to solve the equation of motion for $\phi$ which takes the form 
\be\label{eom:phi}
\ddot\phi+3H\dot\phi+m^{4-n}\phi^{n-1}=0.
\ee
For $n=2$, this is a linear second order differential equation, and its solution and forthcoming physical results can be found, for instance, in \cite{standardclock}. In this paper we mainly focus on the case of $n>2$. In order to solve Eq.~(\ref{eom:phi}), we introduce a new coordinate $x$ and a new variable $f$ as follows 
\be\label{parameterization}
x=a^{-\frac{3(n-2)}{n+2}},\;\;\;\;f=a^{\frac{6}{n+2}}\frac\phi m.%\;\;\;\;\text{for~}n\neq2,
\ee 
Then Eq.~(\ref{eom:phi}) becomes 
\be\label{eom:f-full}
{d^2f\over dx^2}-\frac{2n^2}{(n-2)^2}\frac{f}{x^2}+\frac{(n+2)^2}{9(n-2)^2}\frac{m^2}{H^2}f^{n-1}=0. 
\ee
In some sense, the first term plays a role as ``acceleration". If the last term dominates over the second term, $f$ will oscillate. The minus sign in the second term indicates that it will damp the oscillation. Once the second term dominates over the last term, the field $\phi$ will not oscillate any more. Actually this is just what we expect. In the beginning, the field $\phi$ is heavy and it oscillates around its local minimum. With the expansion of the universe, the amplitude of oscillation damps, and the oscillation ceases when $m_{\rm eff}(\phi_\text{amp})\lesssim H$. Here one can check that the ratio between the last and the second term is nothing but $m_{\rm eff}^2(\phi)/H^2$ up to a numerical factor. In the heavy-field approximation, Eq.~(\ref{eom:f-full}) is simplified to 
\be\label{eom:f}
{d^2f\over dx^2}+\frac{(n+2)^2}{9(n-2)^2}{m^2\over H^2}f^{n-1}\simeq 0. 
\ee 
%And the ``friction'' term can be neglected. 
After lasting around $N_e^\text{heavy}$ e-foldings, the ``friction'' term begins to dominate, and $\phi$ stops oscillating. One thing we need to keep in mind is that as the oscillation of $\phi$ only lasts for a limited period of time for $n>2$, and the oscillatory features in the power spectrum/bispectrum only exists for a specific segment of wave numbers. In this paper we focus on the period when $\phi$ is heavy. 

%%%%%%%%%%%%%%%%%%%%%%%%%%%%%%%%%%%%%%%%%%%%%%%
%\begin{figure}
%\centering
%\includegraphics[width=0.8\textwidth]{phi-t.pdf}
%\caption{The difference between the solution of the e.o.m. \eqref{eom:phi} with $n=2$ (blue) and $n=6$ (red), while $\mu=100$. In $n=6$ case $\phi$ will stop oscillating after a few efolds while in $n=2$ case it will oscillate forever.}
%\end{figure}
%%%%%%%%%%%%%%%%%%%%%%%%%%%%%%%%%%%%%%%%%%%%%%%

\begin{figure}
\centering
\includegraphics[width=0.8\textwidth]{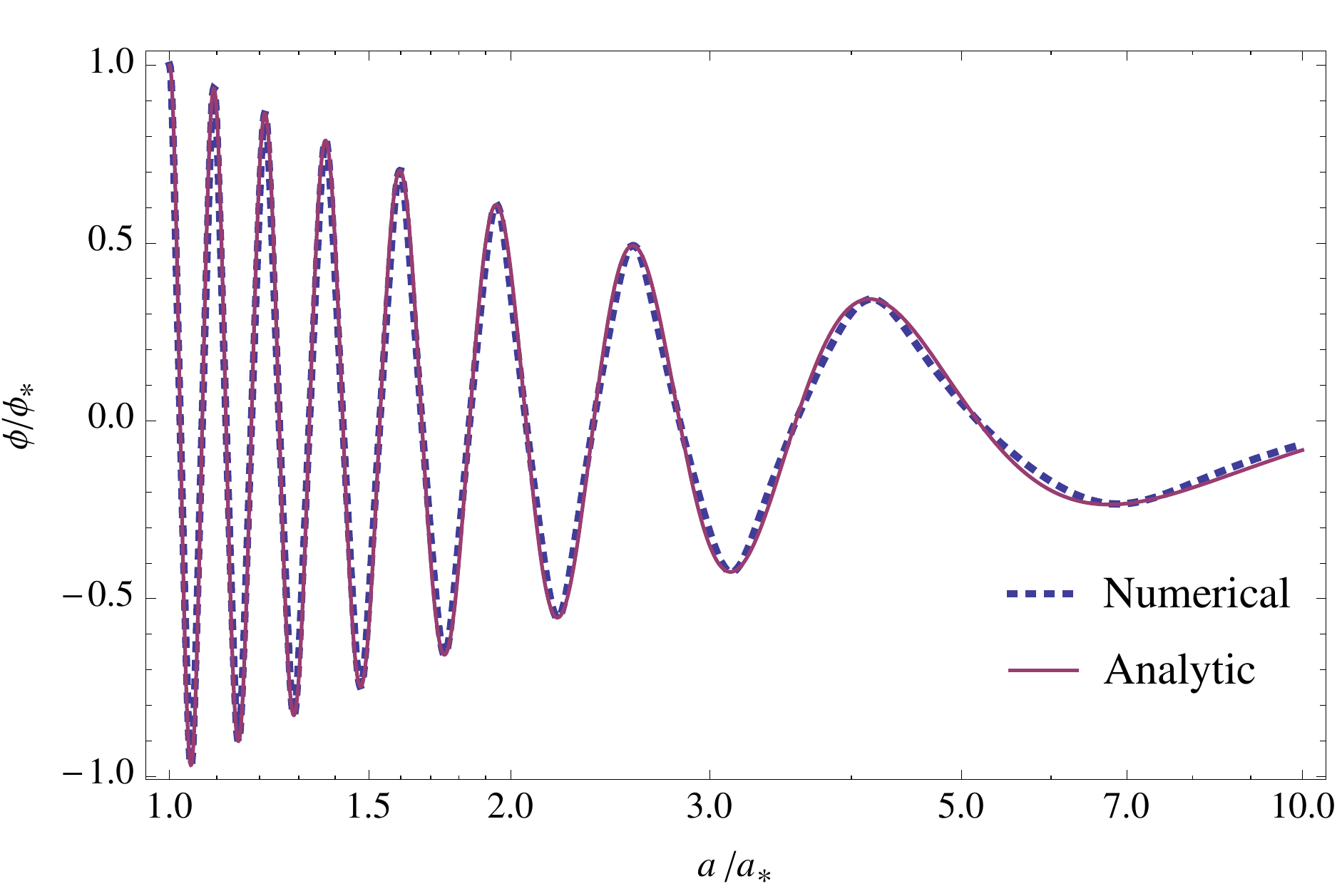}
\caption{The difference between the numerical solution and semi-analytical solution for the equation of motion \eqref{eom:phi} with $n=6$ and $\mu=100$.}
\label{fig:phi-aa}
\end{figure}

The first integral of \eqref{eom:f} gives  
\m\label{integral1}
\({df\over dx}\)^2=C_1-{2(n+2)^2\over 9n(n-2)^2}{m^2\over H^2}f^n. 
\n
%From above we see for a real $df/dx$, $C_1$ should be positive, and
%\be\label{cond:C1}
%\frac{2(n+2)^2}{9n(n-2)^2}\frac{m^2}{H^2}f^{n}\leq C_1.
%\ee
To determine the integral constant $C_1$, we need to connect $df/dx$ with $\dot\phi$, which can be done by recalling the definition of $f$ and $x$,
\be\label{def:g}
\frac{df}{dx}=-\frac{2}{n-2}a^{\frac{3n}{n+2}}\frac\phi m-\frac{n+2}{3n-6}a^{\frac{3n}{n+2}}\frac{\dot\phi}{m H}.%\approx-\frac{n+2}{3n-6}a^{\frac{3n}{n+2}}\frac{\dot\phi}{m H}.
\ee
%The approximation holds when $\phi$ is heavy, i.e. $\dot\phi\gg H\phi$. 
For the initial conditions, we take $\phi(t_*)=\phi_*$, and $\dot\phi(t_*)=0$, which in turn gives
\be\label{def:C1}
C_1=\frac{4}{(n-2)^2}a_*^{\frac{6n}{n+2}}\frac{\phi_*^2}{m^2}+
\frac{2(n+2)^2}{9n(n-2)^2}\frac{m^2}{H^2}f_\ast^{n}\simeq \frac{2(n+2)^2}{9n(n-2)^2}\frac{m^2}{H^2}f_\ast^{n}, 
\ee 
where we consider $\mu\gg 1$ in the last step. Now Eq.~\eqref{integral1} can be written as
\be\label{1stintegral}
\left(\frac{df}{dx}\right)^2\simeq \frac{2(n+2)^2}{9n(n-2)^2}\frac{m^2}{H^2}(f_\ast^n-f^{n}).
\ee
%This, together with \eqref{def:g}, gives
%\be
%\dot\phi^2+\frac{12}{n+2}H\phi\dot\phi+\frac{36}{(n+2)^2}H^2\phi^2+\frac{2}{n}m^{4-n}\phi^n-\frac2nm^{4-n}\phi_\ast^n\left(\frac{a_\ast}{a}\right)^{\frac{6n}{n+2}}=0.
%\ee
%\be\label{def:dotphi}
%\dot\phi=-\frac{6}{n+2}H\phi\pm\sqrt{\frac2nm^{4-n}\phi_\ast^n\left(\frac{a_\ast}{a}\right)^{\frac{6n}{n+2}}-\frac{2}{n}m^{4-n}\phi^n},
%\ee
%which will be used later after we find the approximate solution of $\phi$. The plus/minus sign should be picked in a way that $\dot\phi$ oscillate with half a period retarted than $\phi$. 
%
%By integrating \eqref{1stintegral}, we can have
%However, if we calculate the ratio of the $(m/H)^2f_*^n$ term and the $\phi_*^2/m^2$ term on the right hand side of \eqref{1stIntegral}, we find that it is again proportional to the squared mass parameter $\mu^{2}$. Therefore the $\phi_*^2/m^2$ term in \eqref{1stintegral} can be neglected. After doing so, we can integrate \eqref{1stIntegral} to get% the solution to equation \eqref{eom:f},
The solution of the above equation reads 
\be\label{sol:x}
x=C_2+\frac{f}{\sqrt{C_1}}~{}_2F_1 \left(\frac12,\frac1n;1+\frac1n;\frac{f^n}{f_*^n}
%\frac{2(n+2)^2}{9C_1n(n-2)^2}\left(\frac{m}{H_0}\right)^2f^n
\right).
\ee
where $C_2$ is another integral constant which can be fixed by the initial condition as well. The solution of Eq.~\eqref{eom:f} is the inverse function of $\eqref{sol:x}$, $f(x)$, %They can be fixed by requiring $\phi=\phi_\ast$ and $\dot\phi_\ast=0$ at $a=a_\ast$. %The solution to $f$ is the inverse function of it, with the initial condition of \eqref{def:C1}. The solution
which is a periodic function in $x$-axis. To get its period, we can calculate the the distance on the $x$-axis from $f=0$ to $f_*$, and find the period as follows 
\be
\text{period}=4\frac{f_*}{\sqrt{C_1}}{}_2F_1\left(\frac12,\frac1n;1+\frac1n;1\right)
=a_*^{\frac{6-3n}{n+2}}\frac{12}\mu\sqrt{\frac {n\pi}2}\frac{|n-2|}{n+2}
\frac{\Gamma(1+\frac1n)}{\Gamma(\frac12+\frac1n)}.
%4\frac{{}_2F_1\left(\frac12,\frac1n;1+\frac1n;1\right)}{{}_2F_1\left(\frac12,\frac1n;1+\frac1n;0\right)}=4\frac{\sqrt{\pi}\Gamma \left(1+\frac{1}{n}\right)}{\Gamma \left(\frac{1}{2}+\frac{1}{n}\right)}.
\ee
%We put the determination of the integration constants and the period in the Appendix. 
After obtaining the period of $f(x)$, just for simplicity, we suppose that this periodic function be mimicked by a sinusoidal function, which gives an approximate solution for $\phi(a)$ with a decaying amplitude as follows 
\bea\label{sol:phi}
\phi(a)\simeq \phi_\ast\left(\frac{a_\ast}{a}\right)^{\frac{6}{n+2}}\cos\Xi(a), 
\eea
where 
\bea
\Xi(a)=\sqrt{\frac{\pi}{2n}}\frac{n+2}{3|n-2|}\frac{\Gamma(1/2+1/n)}{\Gamma(1+1/n)}\mu\left[\left(\frac{a_\ast}{a}\right)^{\frac{3(n-2)}{n+2}}-1\right].
\eea
We see that the envelop of the oscillatary amplitude of $\phi$ decreases as that given in \eqref{phiamp}, which is obtained in \cite{Turner:1983he} by a different method. Note that the frequency of the oscillation is proportional to $\mu$ which is much larger than one. Therefore $\phi$ oscillates for many periods within one Hubble time $H^{-1}$, and this justifies our presumption that $\phi$ is heavy. Even though process to get the solution \eqref{sol:phi} seems only valid for $n>2$ apparently, it can however recover the case for $n=2$. Taking $n=2+\Delta$ with the limit of $\Delta\rightarrow 0$, we find 
\be
\phi=\phi_\ast\left(\frac{a_\ast}{a}\right)^{3/2}\cos\left(\mu\ln\frac{a}{a_\ast}\right),
\ee
which is the same as that in~\cite{Kofman:1997yn}. Furthermore, in order to verify that our semi-analytical solution is a quite nice approximate solution, we show the difference between the numerical result and our semi-analytical solution to Eq.~\eqref{eom:f} for a special case $n=6$ in Fig.~\ref{fig:phi-aa}.

Even though our semi-analytical solution is quite close to the numerical solution, taking a derivative to it needs more caution. %if $\dot\phi$ or $\ddot\phi$ appeared in the calculation, we should turn back to the equation of motion \eqref{eom:f} and its first integration for a more accurate expression.  This implies that although \eqref{sol:phi} approximates the solution to \eqref{eom:phi} well, the derivative to it, $\dot\phi$, can not be obtained by simply taking the derivative of $\eqref{sol:phi}$. 
An appropriate way to get $\dot\phi$ is to go back to the first integral \eqref{integral1} and \eqref{def:g} to get the %equation of motion \eqref{eom:f} and its the first integral to find their
algebraic relation with $\phi$ expressed in \eqref{sol:phi}. Another algebraic relation can be found in \eqref{eom:phi} that connects $\ddot\phi$ to $\phi$ and $\dot\phi$. Straightforward algebraic calculations yield the results:% It is thus \eqref{def:dotphi}, which, together with \eqref{sol:phi}, can be written as
%\be\label{sol:dotphi}
%\dot\phi=m^{2-n/2}\phi_\ast^{n/2}\Upsilon(a),
%\ee
%where
%\bea\label{def:Upsilon}
%\Upsilon(a)&=&\mp\sqrt{\frac2n}\left(\frac{a_\ast}{a}\right)^{\frac{3n}{n+2}}\sqrt{1-\cos^n\Phi(a)+\frac{18n}{\xi^2(n+2)^2}}-\frac{6}{n+2}\left(\frac{a_\ast}{a}\right)^{\frac{6}{n+2}}\frac{\cos\Phi(a)}{\xi},\\\label{def:Phi}
%\Phi(a)&=&\sqrt{\frac{\pi}{2n}}\frac{n+2}{3|n-2|}\frac{\Gamma(1/2+1/n)}{\Gamma(1+1/n)}\xi\left[\left(\frac{a_\ast}{a}\right)^{3\frac{n-2}{n+2}}-1\right].
%\eea
%Similarly, to get $\ddot\phi$ we should turn back to the equation of motion \eqref{eom:phi}, and use the result in \eqref{sol:phi} and \eqref{sol:dotphi},
%\be\label{sol:ddotphi}
%\ddot\phi=-m^{4-n}\phi_\ast^{n-1}\left[\left(\frac{a_\ast}{a}\right)^{6\frac{n-1}{n+2}}+\frac3\xi\left(1+\delta\left(\frac{a_\ast}{a}\right)^{\frac{6n}{n+2}}\right)^{1/2}\Upsilon(a)\right].
%\ee
\bea\nonumber
\dot\phi(a)&=&\mp\mu H\phi_\ast\sqrt{\frac2n}\left(\frac{a_\ast}{a}\right)^{\frac{3n}{n+2}}\sqrt{1-\cos^n\Xi(a)+\frac{18n}{\mu^2(n+2)^2}\left(\frac{a}{a_\ast}\right)^{6\frac{n-2}{n+2}}\cos^2\Xi(a)}\\\label{sol:dotphi}
&&-\frac{6}{n+2}\left(\frac{a_\ast}{a}\right)^{\frac{6}{n+2}}H\phi_\ast\cos\Xi(a),\\\label{sol:ddotphi}
\ddot\phi(a)&=&-\mu^2H^2\phi_\ast\left(\frac{a_\ast}{a}\right)^{\frac{6(n-1)}{n+2}}\cos^{n-1}\Xi(a)-3H\dot\phi(a).
\eea
Recall $\mu\gg1$, we see the leading order in \eqref{sol:dotphi} is $(1-\cos^n\Xi)$ in the square root and then $\dot\phi$ is order of $\mu H\phi_\ast$. Similarly, the leading order in \eqref{sol:ddotphi} is the first term which is order of $\mu^2H^2\phi_\ast$. This can also be seen from \eqref{sol:phi}: As the frequency of $\phi(a)$ is roughly $\mu$, every derivative with respect to $t$ will contribute a factor of $\mu H$. This fact is helpful for us to write down the equation of motion for the curvature perturbation at the leading order contributed from the heavy field.

From the Friedmann equation \eqref{Friedmann},
%\be\label{def:H}
%H^2
%=\frac1{3\mpl^2}\left(\frac12\dot\phi_\text{inf}+V_\text{sr}(\phi_\text{inf})+\frac12\dot\phi^2+\frac1nm^{4-n}\phi^n\right).
%\ee
we calculate the corrections from the heavy field $\phi$ to the Hubble parameter,
%\eqref{sol:dotphi},
\bea\nonumber
H^2&=&H_0^2\left\{1+\frac{\epsilon_\text{sr}}{3}-\sqrt{2\epsilon_\text{sr}}\frac{\phi_\text{inf,0}-\phi_\text{inf}}{\mpl}+\frac{\eta_\text{sr}}{2}\frac{(\phi_\text{inf,0}-\phi_\text{inf})^2}{\mpl^2}+\cdots\right.\\\nonumber
&&+\Omega_{\phi*}\left[\left(\frac{a_\ast}{a}\right)^{\frac{6n}{n+2}}\mp\left(\frac{a_\ast}{a}\right)^3\frac{6\sqrt{2n}}{\mu(n+2)}\cos\Xi(a)\sqrt{1-\cos^n\Xi(a)+\frac{18n}{\mu^2(n+2)^2}\left(\frac{a}{a_\ast}\right)^{6\frac{n-2}{n+2}}\cos^2\Xi(a)}\right.\\\label{Hfull}
&&+\left.\left.\frac{36n}{\mu^2(n+2)^2}\left(\frac{a_\ast}{a}\right)^{\frac{12}{n+2}}\cos^2\Xi(a)\right]\right\},
\eea
where
\be
H_0^2=\frac{V_\text{sr}(\phi_\text{inf,0})}{3\mpl^2},
\ee
is the constant part of Hubble parameter. Also, we define the slow-roll parameters for the inflaton $\phi_\text{inf}$,
%while $\dot\phi_\text{inf}$ gives the slow-roll corrections to it. The $\phi$-related terms, however, gives the corrections from the heavy sector, which can be obtained by using \eqref{sol:phi} and 
\be
\epsilon_\text{sr}=\frac{\mpl^2}{2}\left(\frac{V_\text{sr}'(\phi_\text{inf})}{V_\text{sr}(\phi_\text{inf})}\right)^2,
\;\;\;\;
\eta_\text{sr}=\mpl^2\frac{V_\text{sr}''(\phi_\text{inf})}{V_\text{sr}(\phi_\text{inf})},
\;\;\;\;
\xi_\text{sr}^2=\mpl^2\frac{V_\text{sr}'(\phi_\text{inf})V_\text{sr}'''(\phi_\text{inf})}{V_\text{sr}(\phi_\text{inf})^2},
\ee
which are all much smaller than one to sustain inflation. 
%$\Omega_{\phi*}\equiv m^{4-m}\phi_\ast^n/(3n\mpl^2H_0^2)$ is the initial ratio of the energy density of the heavy field to that of the inflaton. 
The corrections from the heavy sector in \eqref{Hfull} are proportional to $\Omega_{\phi*}$ as we expected. Taking the derivative of \eqref{Friedmann} with respect to $t$, and using the definition of the slow-roll parameter, $\epsilon=-\dot H/H^2$, we have
\bea\nonumber
\epsilon&=&\epsilon_\text{sr}+\frac{\dot\phi^2}{2H^2\mpl^2},\\\label{def:epsilon}
&\simeq&\epsilon_\text{sr}+3\Omega_{\phi*}\left(\frac{a_\ast}{a}\right)^{\frac{6n}{n+2}}(1-\cos^n\Xi(a)).
%+\mathcal O(\Omega_{\phi*}^2,\mu^{-2}).
\eea
%In the last step we use the solution \eqref{sol:dotphi}, and only keep the leading order corrections.
Similarly, we have
\bea
\dot\epsilon&\simeq&\dot\epsilon_\text{sr}\pm3\sqrt{\frac n2}\Omega_{\phi*}\mu H\left(\frac{a_\ast}{a}\right)^{\frac{9n-6}{n+2}}\cos^{n-1}\Phi(a)\sqrt{1-\cos^n\Xi(a)},
%+\mathcal O(\Omega_{\phi*}^2,\mu^{-2}),
\\
\ddot\epsilon&\simeq&\ddot\epsilon_\text{sr}-\frac32n\Omega_{\phi*}\mu^2H^2\left(\frac{a_\ast}{a}\right)^{\frac{12n-12}{n+2}}\left[\frac{3n-2}{n}\cos^{2n-2}\Xi(a)-\frac{2n-2}n\cos^{n-2}\Xi(a)\right].
%+\mathcal O(\Omega_{\phi*}^2,\mu^{-2}).
\eea

%Another equation needed is the relation between $a$ and $\tau$. To the leading order we know $a=-(H\tau)^{-1}$ for slow-roll inflation. But now there are some corrections in $H$ and thus
%\be
%\tau=-\int^a_{-\infty}\frac{da}{a^2H}=-\frac1{H_\text{sr}a}\left[1-\delta\frac{n+2}{7n+2}\left(\frac{a_\ast}{a}\right)^{\frac{6n}{n+2}}\right],
%\ee
%whose inverse relation gives
%\be
%a=-\frac{1}{H_\text{sr}\tau}\left[1-\delta\frac{n+2}{7n+2}\left(-k_\ast\tau\right)^{\frac{6n}{n+2}}\right]+\mathcal O(\delta^2).
%\ee
%Here $k_\ast\equiv H_\text{sr}a_\ast$ is the momentum that ``almost'' exits the slow-roll Horizon $H_\text{sr}^{-1}$ at $a_\ast$.

\section{Modulations of the scalar power spectrum}\label{sec:pert}

The evolution of the curvature perturbation in the comoving slice $\mathcal R$ is governed by the equation of motion~\cite{Mukhanov:1990me}, 
\be
u_k''+\left(k^2-\frac{z''}{z}\right)u_k=0,
\ee
where $u_k=z\mathcal R$, $z=\sqrt{2\epsilon}a\mpl$, and the prime denotes the derivative with respect to the comoving time $\tau\equiv\int dt/a$. Using the definition of $z$, the above equation reads
\be\label{eom:uk}
u''_k+\left[k^2-2a^2H^2\left(1-\frac\epsilon2+\frac{3\dot\epsilon}{4H\epsilon}-\frac{\dot\epsilon^2}{4H^2\epsilon^2}+\frac{\ddot\epsilon}{4H^2\epsilon}\right)\right]u_k=0.
\ee
%Before we simplify this equation of motion, we have to carefully compare $\epsilon_\text{sr}$, $\Omega_{\phi*}$ and $\mu^{-1}$. We turn to \eqref{def:epsilon}, 
%\be\label{epsilonLO}
%\epsilon\approx\epsilon_\text{sr}+3\Omega_{\phi*}\left(\frac{a_\ast}{a}\right)^{\frac{6n}{n+2}}(1-\cos^n\Xi(a))+\cdots.\nonumber
%\ee 
In order to keep the leading order of the power spectrum of curvature perturbation the same as that in standard single-field slow-roll inflation, we assume that the slow-roll parameter $\epsilon$ is still dominated by the slow-roll part from the inflaton, namely $\epsilon\simeq \epsilon_\text{sr}$ which implies 
\be\label{cond1}
\Omega_{\phi*}\ll\epsilon_\text{sr}.
\ee
%This means that the (nearly) constant $\epsilon_\text{sr}$ term dominates \eqref{epsilonLO}. 
This condition allows us to simply replace all the $\epsilon$'s in the denominators by $\epsilon_\text{sr}$. %Secondly, we assume
%\be\label{cond2}
%\frac{\delta}{\epsilon_\text{sr}}\xi\gg\eta_\text{sr}.
%\ee
%This ensures that the correction terms will dominate $\eta$ in
Therefore, 
\bea\label{epsilondot}
\frac{\dot\epsilon}{H\epsilon}&=&-2\eta_\text{sr}+4\epsilon_\text{sr}\pm3\sqrt{\frac n2}\frac{\Omega_{\phi*}}{\epsilon_\text{sr}}\mu\left(\frac{a_\ast}{a}\right)^{\frac{9n-6}{n+2}}\cos^{n-1}\Xi(a)\sqrt{1-\cos^n\Xi(a)},\\\nonumber
\frac{\ddot\epsilon}{H^2\epsilon}&=&36\epsilon_\text{sr}^2+4\eta_\text{sr}^2-28\epsilon_\text{sr}\eta_\text{sr}
+2\xi_\text{sr}^2\\\label{epsilonddot}
&&-\frac32n\frac{\Omega_{\phi*}}{\epsilon_\text{sr}}\mu^2\left(\frac{a_\ast}{a}\right)^{\frac{12n-12}{n+2}}\left[\frac{3n-2}n\cos^{2n-2}\Xi(a)-\frac{2n-2}n\cos^{n-2}\Xi(a)\right].
\eea 
The leading correction from the heavy field is the term proportional to $\mu^2$ contained in term of $\ddot \epsilon/(H^2\epsilon)$. The reason is that for the heavy field the derivative with respect to $t$ is order of $\mu H$, where $\mu\gg 1$, and then the term with highest order of derivative makes the main contribution. Here we focus on the modulation of the power spectrum of curvature perturbation from the heavy field, and then the equation of motion of the curvature perturbation reads  
\be\label{eom:uksimplified}
u_k''+\left\{k^2-\frac{2}{\tau^2}\left[1
%-\frac{\epsilon_\text{sr}}{2}+\frac34\eta_\text{sr}
-\frac38n\frac{\Omega_{\phi*}}{\epsilon_\text{sr}}\mu^2\left(-k_\ast\tau\right)^{\frac{12n-12}{n+2}}
\left(\frac{3n-2}{n}\cos^{2n-2}\Xi(\tau)-\frac{2n-2}n\cos^{n-2}\Xi(\tau)\right)\right]\right\}u_k=0, 
\ee
where 
\be
\Xi(\tau)=\sqrt{\frac{\pi}{2n}}\frac{n+2}{3|n-2|}\frac{\Gamma(1/2+1/n)}{\Gamma(1+1/n)}\mu\left[\left(-k_\ast\tau\right)^{3\frac{n-2}{n+2}}-1\right].
\ee 
Assuming that the correction from the heavy field does not break the standard prediction of single-field slow-roll inflation at the leading order, we require 
\be\label{cond2}
\frac{\Omega_{\phi*}}{\epsilon_\text{sr}}\mu^2\ll 1  
\ee 
which is more restricted than that in Eq.~\eqref{cond1}.
%, and \eqref{cond2} alone can be the main assumption of our calculation from now on.
%As we can see from \eqref{cond:Omega}, the correction term in \eqref{eom:uksimplified} is much smaller than the ordinary part identical to that of the single-field inflaiton. So under this condition we can treat the correction term as a perturbation and solve equation 

Here the corrections are treated perturbatively, and then Eq.~\eqref{eom:uksimplified} can be solved by iteration method\textcolor{red}{\footnote{The correction from the slow-roll corrections in the inflaton can be also treated perterbatively, and linearly superposed to the corrections from the heavy field.}}. At the leading order, Eq.~\eqref{eom:uksimplified} reads 
\be
{u_k^{(0)}}''+\left(k^2-\frac2{\tau^2}%\left(1-\frac{\epsilon_\text{sr}}{2}+\frac34\eta_\text{sr}\right)
\right)u_k^{(0)}=0
\ee
whose solutions are 
\be\label{sol:uk0}
u_{k,+}^{(0)}=\frac1{\sqrt{2k}}\left(1-\frac{i}{k\tau}\right)e^{-ik\tau},\;\;\;\;u_{k,-}^{(0)}=u_{k,+}^{(0)\ast},
\ee
where ``*'' denotes the complex conjugate. We use this solution to replace $u_k$ in the correction terms in \eqref{eom:uksimplified}. 
We choose the standard Bunch-Davies vacuum~\cite{Bunch:1978yq} which implies that only the positive-frequency solution $u_{k,+}^{(0)}$ is picked, and then Eq.~\eqref{eom:uksimplified} becomes 
\be\label{inhomo}
u_k''+\left(k^2-\frac2{\tau^2}\right)u_k=\Delta(\tau)u_{k,+}^{(0)},
\ee
where
\be
\Delta(\tau)=-\frac{3}{4}n\frac{\Omega_{\phi*}}{\epsilon_\text{sr}}\mu^2k_\ast^2\left(-k_\ast\tau\right)^{\frac{10n-16}{n+2}}
\left(\frac{3n-2}{n}\cos^{2n-2}\Xi(\tau)-\frac{2n-2}n\cos^{n-2}\Xi(\tau)\right).
\ee
The solution to this inhomogeneous differential equation \eqref{inhomo} is $u_{k,+}^{(0)}+u_k^{(1)}$, where
\bea\nonumber
u_k^{(1)}(\tau)&=&iu_{k,+}^{(0)}(\tau)\int d\tau \left|u_{k,+}^{(0)}(\tau)\right|^2\Delta(\tau)-iu_{k,-}^{(0)}(\tau)\int d\tau \left(u_{k,+}^{(0)}(\tau)\right)^2\Delta(\tau),\\
&=&-\frac34n\frac{\Omega_{\phi*}}{\epsilon_\text{sr}}\mu^2\frac{ik_\ast}{2k}
\left\{\left(1-\frac{i}{k\tau}\right)e^{-ik\tau}%u_{k,+}^{(0)}(\tau)
I_1(k,x)
-\left(1+\frac{i}{k\tau}\right)e^{ik\tau}%-u_{k,-}^{(0)}(\tau)
I_2(k,x)\right\},\label{sol:uk1}
\eea
and 
\bea\label{I1}
I_1&=&\int dx\left(1+\frac{k_\ast^2}{k^2x^2}\right)x^{\frac{10n-16}{n+2}}\left[\frac{3n-2}{n}\cos^{2n-2}\Xi(x)-\frac{2n-2}n\cos^{n-2}\Xi(x)\right],
\\\nonumber
I_2&=&\int dx\left(1-\frac{ik_\ast}{kx}\right)^2x^{\frac{10n-16}{n+2}}e^{2i(k/k_\ast)x}\left[\frac{3n-2}{n}\cos^{2n-2}\Xi(x)-\frac{2n-2}n\cos^{n-2}\Xi(x)\right].\\\label{I2}
\eea
Here $x=-k_\ast\tau$ is defined as a dimensionless integration variable, and
\be\label{Xi-x}
\Xi(x)=\sqrt{\frac{\pi}{2n}}\frac{n+2}{3|n-2|}\frac{\Gamma(1/2+1/n)}{\Gamma(1+1/n)}\mu\left(x^{3\frac{n-2}{n+2}}-1\right).
\ee
%After we get the solution \eqref{sol:uk1} we can calculate 

The primordial power spectrum of the curvature perturbation $\mathcal R$ is %in the comoving gauge: 
\be
\mathcal{P_R}=\frac{k^3}{2\pi^2}\lim_{\tau\rightarrow0}\left|\frac{u_k^{(0)}+u_k^{(1)}}{z}\right|.
\ee
Taking the resulf of \eqref{sol:uk1}, together with the solution in \eqref{sol:uk0}, we have
\be
\mathcal{P}_\mathcal{R}=\mathcal{P}_\mathcal{R}^{(0)}\left\{1+\frac32n\frac{\Omega_{\phi*}}{\epsilon_\text{sr}}\mu^2\frac{k_\ast}{k}\lim_{x\rightarrow0}\textbf{Im}\left[I_1(k,x)+I_2(k,x)\right]\right\},
\ee
where
\be\label{spectrum}
\mathcal{P}_\mathcal{R}^{(0)}=\frac{H_0^2}{8\pi^2\epsilon_\text{sr}\mpl^2}
\ee
is the power spectrum of the single-field inflation, and the corrections $I_1$ and $I_2$ can be expressed by two integrals \eqref{I1} and \eqref{I2}. Note that originally, \eqref{I1} and \eqref{I2} are indefinite integrals. However, adding a constant to either of them is acceptable because the constant can be absorbed into the solution to the homogeneous equation, $u_k^{(0)}$, which in turn is determined by the Bunch-Davies initial conditions. Thus we can replace the indefinite integral in \eqref{I1} and \eqref{I2} by definite integrals from a positive $x$ to $+\infty$. Since there is a $x\rightarrow0$ limit outside, it is also equivalent to set the integral lower limit to $0$. Therefore we are treating two definite integrals on the positive half axis. 

These integrals can be done numerically for any $n\geqslant2$, even for non-interger $n$'s which is common in brane inflation~\cite{Baumann:2014nda}.  For an integer $n$, we can use binomial theorem to expand the power of sinusoidal functions, and integrate $I_1$ and $I_2$ by the stationary phase method term by term. We leave the details of calculation in the Appendix \ref{app:int}, and only write down the final results: 
\bea\nonumber
\frac{\Delta\mathcal{P_R}}{\mathcal{P}_\mathcal{R}^{(0)}}
&=&\frac{\Omega_{\phi*}}{\epsilon_\text{sr}}\frac{3+\text{sgn}(4-n)}{2^n}\sqrt{\frac\pi2\frac{n+2}{|n-4|}}
\left(\sqrt{\frac{2n}\pi}\frac{\Gamma(1+\frac1n)}{\Gamma(\frac12+\frac1n)}\right)^{\frac{21n-30}{4n-16}}
\mu^\half \left(\frac{k}{\mu k_\ast}\right)^{\frac{15n-6}{4n-16}}
\\\nonumber
&&\left\{\left(\frac92n-3\right)\frac1{2^{n}}\sum^{2n-2}_{j=\floor{n}}
\left(\begin{matrix}
       2n-2\\
       j
       \end{matrix}\right)
\left(\frac{1}{j-n+1}\right)^{\frac{21n-30}{4n-16}}\right.\\\nonumber
&&\cdot\left[\sin\Theta^{(1)}_{n,j}
-2\left(\frac{1}{j-n+1}\sqrt{\frac{2n}\pi}\frac{\Gamma(1+\frac1n)}{\Gamma(\frac12+\frac1n)}\right)^{-\frac{n+2}{2n-8}}\mu^{-1}\left(\frac{k}{\mu k_\ast}\right)^{-\frac{3n-6}{2n-8}}\cos\Theta^{(1)}_{n,j}\right.
\\\nonumber
&&-\left.\left(\frac{1}{j-n+1}\sqrt{\frac{2n}\pi}\frac{\Gamma(1+\frac1n)}{\Gamma(\frac12+\frac1n)}\right)^{-\frac{n+2}{n-4}}\mu^{-2}\left(\frac{k}{\mu k_\ast}\right)^{-\frac{3n-6}{n-4}}\sin\Theta^{(1)}_{n,j}\right]
\\\nonumber
&&-(3n-3)\sum^{n-2}_{j=\floor{n/2}}
\left(\begin{matrix}
       n-2\\
       j
       \end{matrix}\right)
\left(\frac{1}{j-n/2+1}\right)^{\frac{21n-30}{4n-16}}
\\\nonumber
&&\cdot\left[\sin\Theta^{(2)}_{n,j}-2\left(\frac{1}{j-n/2+1}\sqrt{\frac{2n}\pi}\frac{\Gamma(1+\frac1n)}{\Gamma(\frac12+\frac1n)}\right)^{-\frac{n+2}{2n-8}}\mu^{-1}\left(\frac{k}{\mu k_\ast}\right)^{-\frac{3n-6}{2n-8}}\cos\Theta^{(2)}_{n,j}\right.
\\\label{result:Pfull}
&&-\left.\left.\left(\frac{1}{j-n/2+1}\sqrt{\frac{2n}\pi}\frac{\Gamma(1+\frac1n)}{\Gamma(\frac12+\frac1n)}\right)^{-\frac{n+2}{n-4}}\mu^{-2}\left(\frac{k}{\mu k_\ast}\right)^{-\frac{3n-6}{n-4}}\sin\Theta^{(2)}_{n,j}\right]\right\}, 
\eea
where $\floor{n}$ is the Gauss floor function that equals to the largest integer not larger than $n$, and 
\bea\nonumber
\Theta^{(1)}_{n,j}(k)&=&\frac{4n-16}{3n-6}\left(\frac1{j-n+1}\sqrt{\frac{2n}\pi}\frac{\Gamma(1+\frac1n)}{\Gamma(\frac12+\frac1n)}\right)^{\frac{n+2}{2n-8}}\mu \left(\frac{k}{\mu k_\ast}\right)^{\frac{3n-6}{2n-8}}\\
&&+\frac{n+2}{3n-6}(2j-2n+2)\sqrt{\frac{\pi}{2n}}\frac{\Gamma(\frac12+\frac1n)}{\Gamma(1+\frac1n)}\mu-\frac\pi4\text{sgn}(n-4),\label{def:Theta1}\\\nonumber
\Theta^{(2)}_{n,j}(k)&=&\frac{4n-16}{3n-6}\left(\frac1{j-n/2+1}\sqrt{\frac{2n}\pi}\frac{\Gamma(1+\frac1n)}{\Gamma(\frac12+\frac1n)}\right)^{\frac{n+2}{2n-8}}\mu \left(\frac{k}{\mu k_\ast}\right)^{\frac{3n-6}{2n-8}}\\
&&+\frac{n+2}{3n-6}(2j-n+2)\sqrt{\frac{\pi}{2n}}\frac{\Gamma(\frac12+\frac1n)}{\Gamma(1+\frac1n)}\mu-\frac\pi4\text{sgn}(n-4).\label{def:Theta2}
\eea
This is our main result. We will discuss this result for different cases individually in the following part of this section. 

\subsection{$n=2$ case}
For $n<4$, only the first terms in each squared brackets in \eqref{result:Pfull} dominate because other terms are all suppressed by some powers of $1/\mu$.
For $n=2$, the second summation series vanish, and the first one has only one term which can be obtained by taking $n=2+\Delta$ with $\Delta\rightarrow 0$. Then we find 
\be
\mathcal{P}_\mathcal{R}=\mathcal{P}_\mathcal{R}^{(0)}
\left\{1+\frac32\sqrt\pi\frac{\Omega_{\phi*}}{\epsilon_\text{sr}}\mu^{1/2}\left(\frac{k}{\mu k_\ast}\right)^{-3}\sin\left[2\mu\ln\left(\frac{k}{\mu k_\ast}\right)+\text{phase}\right]\right\}.\label{result:n=2}
\ee
%This is just the usual $n=2$ result~\cite{standardclock}.(*???*) 
We see that the logarithmic dependence on $k$ in the oscillation is the characteristic feature for a quadratic heavy field $\phi$.

\subsection{$n=3$ case}

When $n=3$, the potential should take the form of $|\phi|^3$. Simplifying Eq.~\eqref{result:Pfull} for $n=3$ gives 
\bea\nonumber
\frac{\Delta\mathcal{P}_\mathcal{R}^{n=3}}{\mathcal{P}_\mathcal{R}^{(0)}}
&=&-\frac{\Omega_{\phi*}}{\epsilon_\text{sr}}\frac{21\sqrt{5\pi}}{2^{11/2}}
\left(\sqrt{\frac{2\pi}3}\frac{\Gamma(5/6)}{\Gamma(3/4)}\right)^{\frac{33}4}\mu^{1/2}\left(\frac{k}{\mu k_\ast}\right)^{-\frac{39}4}\\\label{result:n=3}
&&\cdot\sin\left[\frac43\left(\sqrt{\frac{2\pi}3}\frac{\Gamma(5/6)}{\Gamma(3/4)}\right)^{\frac52}\mu \left(\frac{k}{\mu k_\ast}\right)^{-\frac32}-\frac{10}{3}\sqrt{\frac{2\pi}{3}}\frac{\Gamma(5/6)}{\Gamma(3/4)}\mu-\frac\pi4\right]+\cdots.
\eea
The following terms denoted by dots are suppressed by at least $2^{-33/4}$.

\subsection{$n=4$ case}

Since the argument of the sinusoidal functions in the integrals of $I_1$ and $I_2$ are proportional to $x$, which gives both the two integrals the form of $\int dx x^p e^{i\kappa x}$. However, this integral does not have any stationary point on the positive real axis, and it equals zero if we add a small imaginary part to $x$ when it approaches to $+\infty$. So no oscillatory feature appears in the power spectrum of curvature perturbation for a quartic heavy field. 

%For $n=4$, the result \eqref{result:Pfull} is zero if we expand $n=4+\epsilon$ and take $\epsilon\rightarrow0^+$ limit. An easier way to see this is to go back to \eqref{I1}, \eqref{I2} and \eqref{Xi-x}.  On the contrary, it gives zero if we add a small imaginary part to $x$ when it approches $+\infty$. As a result, no oscillatary feature will appear in the power spectrum of the curvature perturbation for a quartic heavy field.

\subsection{$n>4$ case}

%The amplitude as well as the frequency of the oscillation in \eqref{result:Pfull} are both proportional to $\mu$ to some negative power for $n>4$. This means the amplitude of the oscillation will be suppressed even stronger than $\Omega_{\phi*}/\epsilon_\text{sr}$ which is already much smaller than $1/\mu^2$ according to \eqref{cond2}, and the period will be shorter as $k$ increases. 

In this case, by counting the powers of $\mu$ we see both the last terms in the square brackets dominate. Then we have
\bea\nonumber
\frac{\Delta\mathcal{P}_\mathcal{R}^{n>4}}{\mathcal{P}_\mathcal{R}^{(0)}}
&=&\frac{\Omega_{\phi*}}{\epsilon_\text{sr}}\sqrt{\frac\pi2\frac{n+2}{n-4}}\left(\sqrt{\frac{2n}\pi}\frac{\Gamma(1+1/n)}{\Gamma(1/2+1/n)}\right)^{\frac{17n-38}{4n-16}}\mu^{-{3\over 2}}\left(\frac{k}{\mu k_\ast}\right)^{\frac{3n+18}{4n-16}}\\\nonumber
&&\cdot\left\{-\left(\frac92n-3\right)\frac1{2^{2n-1}}\sum_{j=\floor{n}}^{2n-2}\left(
\begin{matrix}
2n-2\\
j
\end{matrix}\right)\left(\frac1{j-n+1}\right)^{\frac{17n-38}{4n-16}}\sin\Theta^{(1)}_{n,j}(k)\right.\\
&&\left.\quad+\left(3n-3\right)\frac1{2^{n-1}}\sum_{j=\floor{n/2}}^{n-2}\left(
\begin{matrix}
n-2\\
j
\end{matrix}\right)\left(\frac1{j-n/2+1}\right)^{\frac{17n-38}{4n-16}}\sin\Theta_{n,j}^{(2)}(k)\right\}+\cdots,\label{result:P(n>4)}
\eea
In the series, the magnitude decreases as $j$ increases. Therefore the leading term should be the $j=\floor{n}$ and $j=\floor{n/2}$ terms in the first and the second terms in the bracket, and the posterior terms are suppressed at least by $2^{-(17n-38)/(4n-16)}$. We will discuss the cases with an even $n$ or odd $n$ separately.

For an even $n$, $\floor{n/2}=n/2$. At the leading order, since $\Theta^{(1)}_{n,\floor{n}}=\Theta^{(2)}_{n,\floor{n/2}}$, the modulation to the power spectrum of curvature perturbation reads 
\bea\nonumber
\frac{\Delta\mathcal{P}_\mathcal{R}^{n>4,\text{even}}}{\mathcal{P}_\mathcal{R}^{(0)}}
&=&\frac{\Omega_{\phi*}}{\epsilon_\text{sr}}\sqrt{\frac\pi2\frac{n+2}{n-4}}\left(\sqrt{\frac{2n}\pi}\frac{\Gamma(1+1/n)}{\Gamma(1/2+1/n)}\right)^{\frac{17n-38}{4n-16}}\mu^{-{3\over 2}}\left(\frac{k}{\mu k_\ast}\right)^{\frac{3n+18}{4n-16}}\\\nonumber
&&\cdot\left\{\frac{3n-3}{2^{n-1}}\left(
\begin{matrix}
n-2\\
n/2
\end{matrix}\right)
-\frac{\frac92n-3}{2^{2n-1}}\left(
\begin{matrix}
2n-2\\
n
\end{matrix}\right)\right\}\\\nonumber
&&\cdot\sin\left[\frac{4n-16}{3n-6}\left(\sqrt{\frac{2n}\pi}\frac{\Gamma(1+\frac1n)}{\Gamma(\frac12+\frac1n)}\right)^{\frac{n+2}{2n-8}}\mu \left(\frac{k}{\mu k_\ast}\right)^{\frac{3n-6}{2n-8}}\right.
\\
&&\left.\quad\quad+\frac{n+2}{3n-6}\sqrt{\frac{2\pi}{n}}\frac{\Gamma(\frac12+\frac1n)}{\Gamma(1+\frac1n)}\mu-\frac\pi4\right]+\cdots.
\eea
For instance, for $n=6$, we have 
\bea\nonumber
\frac{\Delta\mathcal{P}_\mathcal{R}^{n=6}}{\mathcal{P}_\mathcal{R}^{(0)}}
&=&-\frac{24300}{\sqrt{2}\pi^{7/2}}\frac{\Omega_{\phi*}}{\epsilon_\text{sr}}
\left(\frac{\Gamma(7/6)}{\Gamma(2/3)}\right)^8
\mu^{-{3\over 2}}\left(\frac{k}{\mu k_*}\right)^{9/2}\\
&&\cdot\sin\left[\frac{8}{\pi}\left(\frac{\Gamma(7/6)}{\Gamma(5/6)}\right)^2\mu 
\left(\frac{k}{\mu k_*}\right)^3+\frac23\sqrt{\frac\pi3}\frac{\Gamma(5/6)}{\Gamma(7/6)}\mu-\frac\pi4\right]+\cdots.
\eea
The next-to-leading order terms are suppressed by at least a numerical factor $2^{-8}$.

For an odd $n$, $\floor{n/2}=(n-1)/2$, and then $\Theta^{(1)}_{n,\floor{n}}\neq\Theta^{(2)}_{n,\floor{n/2}}$. However, an additional $2^{(17n-36)/(4n-16)}$ factor enhances the term involving $\Theta^{(2)}_{n,n/2-1/2}$ compared to the term $\Theta^{(1)}_{n,n}$, and therefore the modulation at the leading order is given by 
\bea\nonumber
\frac{\Delta\mathcal{P}_\mathcal{R}^{n>4,\text{odd}}}{\mathcal{P}_\mathcal{R}^{(0)}}
&=&\frac{\Omega_{\phi*}}{\epsilon_\text{sr}}\frac{3(n-1)}{2^{n-1}}\sqrt{\frac\pi2\frac{n+2}{n-4}}\left(2\sqrt{\frac{2n}\pi}\frac{\Gamma(1+1/n)}{\Gamma(1/2+1/n)}\right)^{\frac{17n-38}{4n-16}}\mu^{-{3\over 2}}\left(\frac{k}{\mu k_\ast}\right)^{\frac{3n+18}{4n-16}}\\
&&\left(
\begin{matrix}
n-2\\
\frac n2-\frac12
\end{matrix}\right)
\sin\left[\frac{4n-16}{3n-6}\left(2\sqrt{\frac{2n}\pi}\frac{\Gamma(1+\frac1n)}{\Gamma(\frac12+\frac1n)}\right)^{\frac{n+2}{2n-8}}\mu \left(\frac{k}{\mu k_\ast}\right)^{\frac{3n-6}{2n-8}}\right.\nonumber\\
&&\left.\quad\quad\quad\quad\quad\quad
+\frac{n+2}{3n-6}\sqrt{\frac{\pi}{2n}}\frac{\Gamma(\frac12+\frac1n)}{\Gamma(1+\frac1n)}\mu-\frac\pi4\right]+\cdots.
\eea

Omitting the numerical factor, we find that the modulations of power spectrum of curvature perturbation for $n>4$ has a universal form as follows 
%If we omit the complex numerical factor and only focus on the behavior of $\mu$ and $k$ thus to keep a clearer form for the possible observational search for the signal of the feature, there is a simplfied universal formula for both even and odd $n$'s:
%%%%%%%%%%%%%%%%%%%%%%%%%%%%%%%%%%%%%%%%%%
%%%%%%%				Alternative expression					%%%%%%
%%%%%%%%%%%%%%%%%%%%%%%%%%%%%%%%%%%%%%%%%%
\iffalse
\be\label{result:Psim(n>4)}
\frac{\Delta\mathcal{P}_\mathcal{R}^{n>4}}{\mathcal{P}_\mathcal{R}^{(0)}}
\propto\frac{\Omega_{\phi*}}{\epsilon_\text{sr}}
\left(\frac1\mu\right)^{\frac52s-\frac32}\left(\frac{k}{k_*}\right)^{\frac52s-3}
\sin\left[\left(\frac{C(n)}\mu\right)^{s-1}\left(\frac{k}{k_\ast}\right)^s+\text{phase}\right],
\ee
\fi
%%%%%%%%%%%%%%%%%%%%%%%%%%%%%%%%%%%%%%%%%%
%%%%%%%				Alternative expression					%%%%%%
%%%%%%%%%%%%%%%%%%%%%%%%%%%%%%%%%%%%%%%%%%
\be\label{result:Psim(n>4)}
\frac{\Delta\mathcal{P}_\mathcal{R}^{n>4}}{\mathcal{P}_\mathcal{R}^{(0)}}
\propto\frac{\Omega_{\phi*}}{\epsilon_\text{sr}}\left(\frac{k}{\mu k_*}\right)^{\frac52s-3}\mu^{-3/2}
\sin\left[{C(s)\over s}\mu \left(\frac{k}{\mu k_\ast}\right)^s+\text{phase}\right],
\ee
where ``phase'' is the phase term independent on $k$, $s=(3n-6)/(2n-8)$,
%as is given in Eq.~\eqref{sn}
 and $C(s)$ is a function of $s$ with non-zero value at $s=0$. Here we separate $s$ in the denominator from the other part of the argument in the sinusoidal function because we would like to write explicitly the first order pole at $s=0$ when $n\rightarrow 2$. For $n>4$, we have $s>3/2$. 

From Eq.~\eqref{result:Psim(n>4)}, we see that the argument in the sinusoidal functions is proportional to a positive power of $k$. It indicates that the oscillation period (measured in $k$) will become shorter as $k$ increases. And the amplitude of the correction increases as $k$ increases since it is proportional to a positive power of $k$ for $n>4$.
It is worth noting that even though the amplitude of the oscillation grows up as $k$ increases, it will not break the validity of the perturbation theory because the oscillation lasts a finite number of e-folds. Recalling Eq.~\eqref{efoldheavy}, we have a lower limit and an upper limit on the wavenumber $k$ in the oscillation stage, namely 
\be
\mu^{-1}<\frac{k}{\mu k_\ast}<\mu^{-1/s}.
\ee
Considering $s>3/2$ for $n>4$, we can see that the prefactor of the correction in \eqref{result:Psim(n>4)} is
\be
\frac{\Omega_{\phi*}}{\epsilon_\text{sr}}\left(\frac{k}{\mu k_*}\right)^{\frac52s-3}\mu^{-3/2}
<\frac{\Omega_{\phi*}}{\epsilon_\text{sr}}\mu^{-2}\ll\mu^{-4},
\ee
where Eq.~\eqref{cond2} is used in the last step. %Therefore the perturbation theory is safe throughout the era when $\phi$ is heavy.

%The comoving wavenumber $k$ always appears as a combination of $k/(\mu k_\ast)$ in both \eqref{result:n=2} and \eqref{result:Psim(n>4)}. This means that there is resonance when the mode $k$ crosses the ``mass horizon'' which is defined by $k=\mu H a$. This resonance on the mass-horizon-crossing scale is already observed in \cite{Chen:2008wn,standardclock}, only for the $n=2$ case. Now we can read from \eqref{result:Psim(n>4)} (and \eqref{result:Pfull} too without explicitly written here) the resonance on the mass-horizon-crossing is an ubiquitous property for arbitrary monimial potentials.

\section{Summary and Discusion}\label{sec:con}

In this paper we compute the modulation of the power spectrum of curvature perturbation from a heavy field with a monomial potential $\sim \phi^n$ and find that a power-law modulation generically appears. As the perturbation mode $k$ increases, the amplitude of the sinusoidal modulation decreases  for $2\leq n<4$, while increases for $n>4$. For $n=4$ there is no oscillating modulation. For interger $n$'s, we have analytical results.

In \cite{standardclock} the authors proposed that the oscillations of a heavy spectator field with a potential $m^2\phi^2/2$ can be taken as a good candidate for the physical clock in the early universe. Such oscillations modulate the power spectrum of curvature perturbation as follows 
\be\label{standardclock}
\frac{\Delta\mathcal{P}_\mathcal{R}^\text{s.c.}}{\mathcal{P}_\mathcal{R}^{(0)}}
\sim \left(\frac{k}{k_r}\right)^{\frac5{2}q-3}\sin\left[\frac{1}{q^2-q}2\mu\left(\frac{2k}{k_r}\right)^{q}+\text{phase}\right]. 
\ee
It is interesting that the above modulation looks the same as Eq.~\eqref{result:Psim(n>4)}. In \cite{standardclock} the authors concluded that $q<1$ for inflation and $q>1$ for a contracting phase in the early universe. However, we provide a counter example: $q\geq 3/2$ can be obtained for inflation in which the power spectrum of curvature perturbation is modulated by the oscillations of a heavy field with potential $\sim \phi^n$ ($n>4$).
Therefore, if we find some signal featured by the above formula in the future, whether it comes from the heavy field with an arbitrary monomial potential or just from the quadratic standard clock picture in a contracting phase is indistinguishable. 

Actually we believe that in principle it is impossible to distinguish the inflationary expansion from the contracting phase by concerning the correlation functions of curvature perturbation. Since $H\equiv d\ln a/dt>0$ for the expanding phase and $H<0$ for the contracting phase, we can distinguish the expanding phase from the contracting phase by measuring the sign of $H$. However, in general, the $n$-point correlation function of curvature perturbation takes the form 
\m
\langle \zeta_{{\bf k}_1} \zeta_{{\bf k}_2}\cdots \zeta_{{\bf k}_n}\rangle \propto P_\zeta^{n-1},
\n
where $P_\zeta\propto H^2$. Therefore it is impossible to determined the sign of $H$ by measuring the the $n$-point correlation function of curvature perturbation.

\vspace{5mm}
\noindent {\bf Acknowledgments} \\
We would like to thank Y. Wang for useful conversation.
This work is supported by Top-Notch Young Talents Program of China, grants from NSFC (grant NO. 11322545, 11335012 and 11575271), and Key Research Program of Frontier Sciences, CAS.

\appendix

\section{Some integrals}\label{app:int}

Integral \eqref{I1} is like 
\be\label{app1}
\int^\infty_0 dx~x^ae^{ibx},
\ee
after we expand the cosine function into exponentials. The integrand of \eqref{app1} oscillates with the same period severely, therefore we can add a small imaginary part to $x$ and find that the integral equals to zero on the full positive $x$-axis.

Integral \eqref{I2} has a templet of
\be\label{In}
I(n,l,b)=\int dx~x^{\frac{10n-16}{n+2}-l} e^{2i\kappa x}\cos^b\Xi(x),
\ee
where $l=0,1,2$ correspond to three terms in the bracket in \eqref{I2}, $b=n-2,2n-2$ correspond to two different types of cosine functions in \eqref{I2}, and $\kappa\equiv k/k_\ast$. We express the cosine function as a linear combination of exponential functions, then use the binomial series to expand it,
\be\label{temp0}
\cos^b\Xi=\frac1{2^b}\left(e^{i\Xi}+e^{-i\Xi}\right)^b=\frac{1}{2^b}e^{ib\Xi}\left(1+e^{-2i\Xi}\right)^b
=\frac1{2^b}e^{ib\Xi}\sum^\infty_{j=0}\left(
\begin{matrix}
b\\
j
\end{matrix}\right)e^{-2ji\Xi},
\ee
where
\be
\left(
\begin{matrix}
b\\
j
\end{matrix}\right)=\frac{b(b-1)(b-2)\cdots(b-j+1)}{j!}
\ee
are the binomial coefficients. If $b$ is an integer, \eqref{temp0} will be truncated at $(b+1)$-th term because all the terms later will be multipled by 0. If $b$ is non-integer, \eqref{temp0} will become infinite series.

Using \eqref{temp0}, we can represent \eqref{In} as
\be\label{temp11}
I(n,l,b)=\frac{1}{2^b}\sum^\infty_{j=0}\left(
\begin{matrix}
b\\
j
\end{matrix}\right)\int dx~x^{\frac{10n-16}{n+2}-l}e^{i\Lambda\omega(x)},
\ee
%while for our case $a=0,-1,-2$, $b=n-2$ or $2n-2$, and $\kappa=k/k_\ast$.
%If $b$ is an even integer, we have
%\be
%\cos^b\Phi=\frac1{2^{-b}}\left(
%\begin{matrix}
%b\\
%b/2
%\end{matrix}\right)
%+\frac{1}{2^{-b}}\sum^{b/2-1}_{j=0}\left(
%\begin{matrix}
%b\\
%j
%\end{matrix}
%\right)\left(e^{i(b-2j)\Phi}+e^{-i(b-2j)\Phi}\right).
%\ee
%And
%\be
%I_n(a,b)=\sum^{b/2-1}_{j=0}\left(
%\begin{matrix}
%b\\
%j
%\end{matrix}
%\right)J_n^+(a,b,j)+\sum^{b/2-1}_{j=0}\left(
%\begin{matrix}
%b\\
%j
%\end{matrix}
%\right)J_-(a,b,j)+\int dz z^a e^{2i\kappa z}.
%\ee
%The last term is oscillating rapidly thus contributes zero. 
%The $J$ integrals are
%\be
%J_n^\pm(a,b,j)=\int dz~z^a\exp\left[2i\kappa z\pm i(b-2j)A_n\xi z^{3\frac{n-2}{n+2}}\right].
%\ee
%If $b$ is an odd integer, we have
%\be
%\cos^b\Phi=
%\frac{1}{2^{-b}}\sum^{(b-1)/2}_{j=0}\left(
%\begin{matrix}
%b\\
%j
%\end{matrix}
%\right)\left(e^{i(b-2j)\Phi}+e^{-i(b-2j)\Phi}\right).
%\ee
%And
%\be
%I_n(a,b)=\sum^{(b-1)/2}_{j=0}\left(
%\begin{matrix}
%b\\
%j
%\end{matrix}
%\right)J_n^+(a,b,j)+\sum^{(b-1)/2}_{j=0}\left(
%\begin{matrix}
%b\\
%j
%\end{matrix}
%\right)J_-(a,b,j).
%\ee
where
\bea
\Lambda&\equiv&(2j-b)\sqrt{\frac{\pi}{2n}}\frac{\Gamma(1/2+1/n)}{\Gamma(1+1/n)}\mu,\\
\omega(x)&\equiv&\frac{2\kappa}\Lambda x-\frac{n+2}{3(n-2)}\left(x^{\frac{3(n-2)}{n+2}}-1\right).
\eea
Here $\Lambda\gg1$ because it is just $\mu$ multipled by a pure number, and $i\Lambda\omega(x)$ is therefore a rapidly oscillating phase factor. This integral can be done by the stationary phase method. We will mostly follow the discussions in~\cite{stationarypoint}. Here $\omega'(x)=0$ gives the stationary point. If $b>2j$, the place with the stationary phase will be on the negative axis, which means an integral from 0 to $+\infty$ gives zero. If $b=2j$, $\omega(x)\propto x$, and the integral contributes to zero too, just as in the $n=4$ case. Only for $b<2j$ there is a positive stationary point lays in
\be
x_\text{sp}=\left(\frac{2\kappa}{\Lambda}\right)^{\frac{n+2}{2(n-4)}}.
\ee
This gives the result
\bea\nonumber
&&\int^\infty_0 dx~x^{\frac{10n-16}{n+2}-l}e^{i\Lambda\omega(x)}\\\nonumber
&\approx&x_\text{sp}^{\frac{10n-16}{n+2}-l}
e^{i\Lambda\omega(x_\text{sp})}\int^\infty_0e^{i\frac\Lambda2\omega''(x_\text{sp})(x-x_\text{sp})^2},\\\label{integralintermediate}
&=&\sqrt{\frac{\pi}{2\Lambda\left|\omega''(x_\text{sp})\right|}}x_\text{sp}^{\frac{10n-16}{n+2}-l}
e^{i\Lambda\omega(x_\text{sp})+i\frac\pi4\text{sgn}(\omega''(x_\text{sp}))}
\left[1-\text{erf}\left(-x_\text{sp}\sqrt{-i\frac\Lambda2\omega''(x_\text{sp})}\right)\right], 
\eea
where 
\bea
\omega(x_\text{sp})&=&\frac{2(n-4)}{3(n-2)}\left(\frac{2\kappa}{\Lambda}\right)^{\frac{3n-6}{2(n-4)}}+\frac{n+2}{3(n-2)},\\
\omega''(x_\text{sp})&=&-\frac{2(n-4)}{n+2}\left(\frac{2\kappa}{\Lambda}\right)^{\frac{n-10}{2(n-4)}},
\eea
and $\text{erf}$ is the error function defined by $\text{erf}(z)=\frac{2}{\sqrt\pi}\int^z_0e^{-t^2}dt$. We notice that both $\omega(x_\text{sp})$ and $\omega''(x_\text{sp})$ are equal to zero for $n=4$. It indicates that the integral in Eq.~\eqref{In} equals zero for $n=4$. To step further let us focus on the argument of the error function, which gives
\be\label{erfarg}
-x_\text{sp}\sqrt{-i\frac\Lambda2\omega''(x_\text{sp})}
=-e^{i\text{sgn}(n-4)\pi/4}\sqrt{\frac{|n-4|}{n+2}}\sqrt{\frac{k_\ast}{2k}}
\left(\frac{2}{2j-b}\sqrt{\frac{2n}{\pi}}\frac{\Gamma(1+1/n)}{\Gamma(1/2+1/n)}\mu^{-1}\right)^{\frac{n+2}{4n-16}}.
\ee
Forget the numerical factors, this argument is roughly $\mu^{(n+2)/(16-4n)}$, which is much larger than 1 for $n<4$, and much smaller than 1 for $n>4$. Therefore, taking into account the imaginary part, and expand the error function at large $\mu$ limit for $n<4$ and at small $\mu$ limit for $n>4$ respectively, we have
\bea
\text{erf}(-e^{-i\pi/4}\#\mu^{\frac{n+2}{16-4n}})
&=&-1+\frac{\mu^{-\frac{n+2}{16-4n}}}{\sqrt\pi}e^{i\#^2\mu^{\frac{n+2}{8-2n}}+i\pi/4}+\cdots,
\;\;\;\;\text{for }n<4;\\
\text{erf}(-e^{i\pi/4}\#\mu^{\frac{n+2}{16-4n}})
&=&-\frac{2}{\sqrt\pi}\#\mu^{-\frac{n+2}{4n-16}}e^{i\pi/4}+\cdots,
\;\;\;\;\text{for }n>4.
\eea
 Here $\#$ represents the insignificent numerical factor in \eqref{erfarg} that we would like not to write down explicitly. If we only forcus on the leading order, the squared bracket in \eqref{integralintermediate} is
\be
\left[1-\text{erf}\left(-x_\text{sp}\sqrt{-i\frac\Lambda2\omega''(x_\text{sp})}\right)\right]
=\left\{
          \begin{matrix}
          \;2, \;\;\;\;\text{for }n<4;\\
          {}\\
          \;1, \;\;\;\;\text{for }n>4,
           \end{matrix}\right.
=\frac{3+\text{sgn}(4-n)}{2}.
\ee
And this gives us the result for integral $I(n,l,b)$ for $n\neq 4$, 
\bea\nonumber
I(n,l,b)&\approx&\frac{3+\text{sgn}(4-n)}{2^{b+2}}\sqrt{\frac{\pi}{2}\frac{n+2}{|n-4|}}
\left(\frac{k}{\mu k_\ast}\right)^{\frac{19n-22}{4n-16}}\mu^{-1/2}
\\\nonumber
&&\sum^\infty_{j>b/2	}\left(
\begin{matrix}
b\\
j
\end{matrix}\right)
\left(\frac{2}{2j-b}\sqrt{\frac{2n}{\pi}}\frac{\Gamma(1+\frac1n)}{\Gamma(\frac12+\frac1n)}\right)^{\frac{21n-30}{4n-16}}\left(\frac{2}{2j-b}\sqrt{\frac{2n}{\pi}}\frac{\Gamma(1+\frac1n)}{\Gamma(\frac12+\frac1n)}\cdot\frac{k}{\mu k_\ast}\right)^{-l\frac{n+2}{2n-8}}\\\nonumber
&&\cdot\exp\left[i\frac43\frac{n-4}{n-2}\left(\frac{2}{2j-b}\sqrt{\frac{2n}{\pi}}\frac{\Gamma(1+\frac1n)}{\Gamma(\frac12+\frac1n)}\right)^{\frac{n+2}{2n-8}}\left(\frac{k}{\mu k_\ast}\right)^{\frac{3n-6}{2n-8}}\mu\right.\\
&&\left.\quad\quad+i\frac{n+2}{3n-6}(2j-b)\sqrt{\frac{\pi}{2n}}\frac{\Gamma(1/2+1/n)}{\Gamma(1+1/n)}\mu+i\frac\pi4\text{sgn}(4-n)\right].\label{sol:J}
\eea

\end{document}